\documentclass{ws-procs10x7}

\usepackage{epsfig}

\def\as{\alpha_s}

\def\qq{q\bar{q}}
\def\ee{e^+e^-}
\def\Kout{K_{\rm out}}
\def\cF{{\cal F}}

\def\VEV#1{\left\langle#1\right\rangle}

\def\be{\beta}

\def\out{{\rm out}}
\def\Eout{E_{\rm out}}
\def\tin{\theta_{\mbox{\scriptsize in}}}

\begin{document}

\title{ Jet-shape observables}

\author{Giuseppe Marchesini}

\address{Dipartimento di fisica, Universit\`a di Milano-Bicocca and\\
INFN, Sezione di Milano, Italy}

\twocolumn[\maketitle\abstract{Studies of jet-shape observables in
  hard processes are summarized together with future developments}]

\section{Motivations for jet-shape study}

Jet-shape studies provide information on QCD dynamics, are essential
tools in analysis of (expected) ``new physics'' events and are needed
for theoretical improvements and tests for Monte Carlo simulations.
Jets-shape measurement involves the bulk of the events.  They provide
$30\%$ of the entries in the determination\cite{SigBet} of $\as$.

Jets are present in all hard events in all high energy processes
($\ee$, DIS and hadron-hadron collisions). Their characteristic
features can be revealed by different jet-shape observables. For
instance, thrust and broadening
$$
B=\sum_h\frac{p_{ht}}{Q}\,,\quad
\tau\!=\!1\!\!-\!\!T\!=\!\sum_h\frac{p_{ht}\,e^{-|\eta_h|}}{Q}\,,
$$
($p_h$ and $\eta_h$ the hadron transverse momentum and rapidity with
respect to jet axis) probe radiation in different rapidity regions.
Other observables, such as
$V\!=\!(1\!-\!T)\,,B\,,C\,,\rho\,,D\,,y_{ij}\,,\Kout \cdots$, probe
various other features.
The jet-shape distributions
$$
\Sigma(V)=\sum_n\int\frac{d\sigma_n}{\sigma_{\rm tot}}\,
\Theta\!\left(V\!-\!\sum_hv_h\right)\,,
$$
are (in general) collinear safe (for $\vec{p}_i,\vec{p}_j$ parallel
one may replace the two hadrons with a single one with
$\vec{p}_i\!+\!\vec{p}_j$) and IR safe (for $p_i \ll p_j$ one may
neglect $p_i$). This implies that all perturbative coefficients of
$\Sigma(V)$ are finite and one may assume parton-flow $\simeq$
hadron-flow.

In the following I discuss the status of the calculations. First I
discuss ``global'' observables (all emitted hadrons are contributing).
Then I discuss the ``non-global'' ones (only hadrons in certain
regions of phase space are contributing, e.g. rapidity cut in
hadron-hadron collisions).  In the first case the relevant
configurations correspond to hadrons within the jets (then each jet
contributes ``independently'' with a Sudakov factor).  In the second
case additional logarithmic terms are contributing which come from
soft gluon emitted away from the jet region (correlations among jets
are generated).

\section{Global observables}
The construction of reliables predictions for $\Sigma(V)$ involves
perturbative and non-perturbative aspects. Before discussing specific
observables I summarize the needed computations\\
1) exact LO and NLO computations (needed for finite $V$)
$$
\Sigma_{\rm PT}(V)\!= \!a(V)\,\as\!+\!b(V)\,\as^2\!\cdots
$$
2) DL and SL resummations (needed for large $L\!=\!\ln V^{-1}$)
$$
\ln \Sigma(V)\!=\!\sum_{n=1}^\infty\!\Big\{\!d_n\,\as^nL^{n+1}\!+\!
s_n\,\as^nL^{n}\cdots\!\Big\}
$$
At least $d_n$ (collinear+IR) and $s_n$ (collinear or IR) terms
need to be resummed (using factorization of collinear and IR
contributions and factorization of observable constraints via Mellin
and/or Fourier transforms);\\
3) matching of exact computations (for finite $V$) and
resummations (for small $V$);\\
4) power corrections originating from the running coupling at any
scale ($0\!<\!k_t\!<\!Q$), thus involving hadron scales (at virtual level).
This leads to non-convergence of PT expansion. For the average value,
for instance,
$$
\VEV{V}_{\rm PT}\!
\sim \!\int^Q_{ 0}\!\!\!\!\!\as(k_t)\frac{dk_t}{Q}\!\sim\!\!{
\sum_n\!  n!\!\left(\!\frac{\be_0}{4\pi}\right)^n\!\!\!\!\as^n}\,.
$$
Without non-perturbative information, at present, the
non-convergence is cured by prescriptions. Here are some: 
i) introduce\cite{DMW} a non-perturbative parameter
$\alpha_0(\mu_I)=\int_0^{\mu_I}\frac{dk_t}{\mu_I}\as(k_t)$ so that
$$
\Sigma(V)=\Sigma_{\rm PT}(V+{ \Delta V})\,,
$$
with $\Delta V$ given by $\alpha_0(\mu_I)$ with computable
coefficient. One has: $n!$-cancellation, $\mu_I$-independence, 
universality;
ii) introduce a shape function\cite{KS}$^,$\cite{GR} to modulate
the radiator at large distance to account for higher ${ 1/Q}$ powers;
iii) improve renormalization group implementation to
reduced power corrections\cite{DM}.

The simplest cases are in $\ee$ with mostly two jet-events.  The
difficulty increases when more than two jets are involved. Here one
has different partonic channel of the hard jets and one can explore in
large contexts different prescriptions for non-perturbative
corrections. Interesting cases are: 1) ``three-jet observables'' in
$\ee$ such\cite{3jet-ee} as $D$-parameter and $\Kout$
(out-of-event-plane momentum) in events with finite $1\!-\!T$; 2)
``three-jet observables'' in DIS such\cite{3jet-DIS} as azimuthal
correlation and $\Kout$-distribution in large $P_t$ events; 3)
hadron-hadron collisions. Except for special cases\cite{3jet-hh} in
which only three QCD jets are involved, typically one has four jets
(two incoming and two outgoing ones). In these cases one encounters
new colour algebra features. With two ($T_1+T_2=0$) or three
($T_1+T_2+T_3=0$) colour charges all products of primary parton colour
matrices $T_iT_j$ are proportional to the identity. This is not any
more true for four ($T_1+T_2+T_3+T_4=0$) or more primary parton
processes.  As a result in hadron-hadron dijet events, there are
peculiar colour correlation\cite{BS} among the four jets.

In conclusion, there is a large variety of jet-shape observables, they
are very informative and characteristic of QCD, they enter all hard
processes with features which are universal (QCD factorization).
However they are difficult to measure (requiring a full knowledge of
the produced hadrons) and to compute. Publications show that, on
average, there is one paper per observable in a single process. Since
by now, in principle, the computation procedure is known, one could
build up an automated calculation

\section{Automated resummation}
A.Banfi, G.Salam and G.Zanderighi have constructed a numerical
program\cite{BSZ} to perform the automated resummation for any global
jet-shape (present or future) distribution and for all processes.
Here is the basic steps performed numerically. First the observable
under consideration is analysed to check for collinear and IR
regularity and globalness. Then one starts from the general factorized
formula. For the most complex hard process, hadron-hadron collisions,
one has
$$
\Sigma_{h_1h_2}(V)=(p^a_{h_1}\cdot p_{h_2}^b)\times M^2_{ab\to
  cd} \times \tilde\Sigma_{ab\to cd}(V)\,,
$$
with $p_h^a$ density of parton $a$, $M^2$ elementary hard
distribution and $\tilde\Sigma(V)$ the radiation formula for the
corresponding observable which involves DL and SL resummation
$$
{ \tilde \Sigma_{\rm res.}(V)=e^{-R(V)}\cdot \cF(\as\ln V)}\,.
$$
Here $R(V)$ is the known (universal) DL-radiator and $\cF(V)$ the
SL-function computed by numerical simulation (for hadron-hadron
collision this requires also the colour correlations among the four
jets\cite{BS}). To match with LO (and NLO) exact results one computes
the coefficient function $C(\as,V)$
$$
{ \tilde \Sigma_{\rm PT}(V)=C(\as,V)\cdot\tilde \Sigma_{\rm res.}(V)}\,.
$$
Finally, power corrections are included\cite{DMW} by
$$
 \tilde \Sigma(V)=\tilde\Sigma_{\rm PT}(V+\Delta V)\,.
$$

\section{New entry: non-global logs}
In hadron-hadron collisions, due to the presence of rapidity cut,
jet-shape observables take contributions only from part of phase space
(non-global observables).  Also many DIS observables\cite{DS-DIS} are
of this type.  Here cancellation\cite{DS} of collinear and IR
singularities between real and virtual contributions is more complex
than for global observables.  There are additional SL contributions
originating\cite{DS}$^,$\cite{BMS}$^,$\cite{BKS}$^,$\cite{YP} from
soft large angle gluons generating correlations among jets not present
in global observables.  Indeed for global observables, due to
dominance of collinear configurations, the radiation of each jet
contributes (essentially) in an independent way.

Consider $\Sigma_{\ee}(Q,\Eout,\tin)$, the distribution of energy
deposited away from a cone $\tin$ around the thrust axis in $\ee$
\begin{figure}[ht]
    \epsfig{file=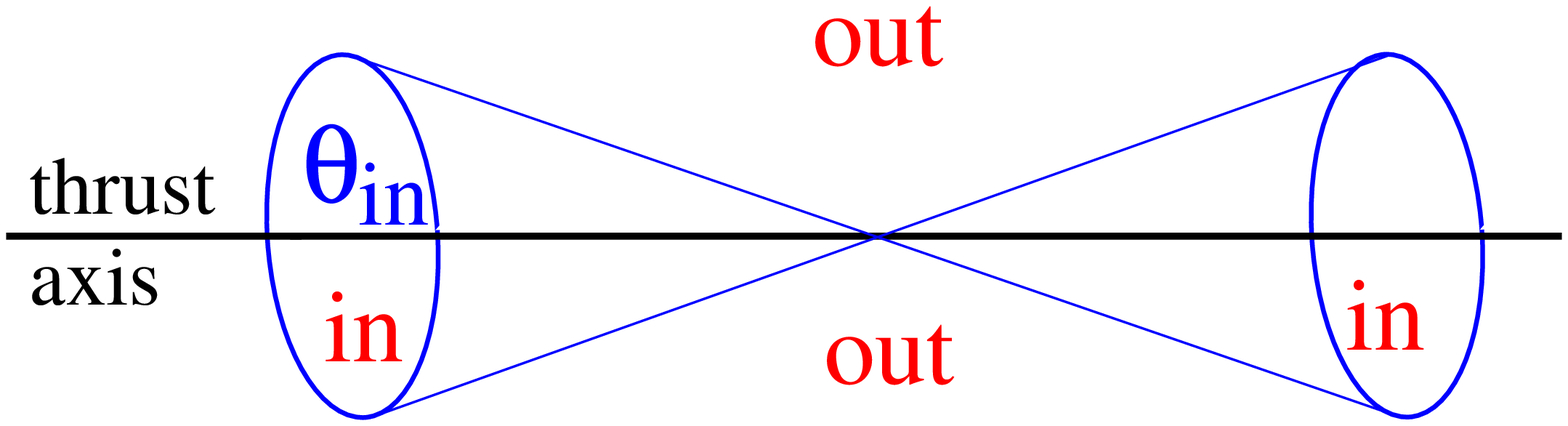,width=0.45\textwidth}
\end{figure}
$$
\Sigma_{\ee}\!=\!\sum_n\!\int\frac{d\sigma_n}{\sigma_T}\>
\Theta\!\left(\!\Eout\!-\!\sum_{  \out}k_{ti}  \right)
$$
There are two PT components. The first takes into account soft
gluons directly emitted in region ``out'' by the primary
quark-antiquark along the thrust axis. This is the usual
bremsstrahlung component (present in global observables) and its
resummation leads to the Sudakov factor $S_{\qq}$ which is SL since
soft gluons are emitted at large angle.  The second component, present
only in non-global observables, takes contributions of soft gluons
which, before entering the observation region ``out'', undergo
branchings within the cone $\tin$ around the jet.  $C_{\qq}$ is the
residue of partial cancellations of collinear and IR logarithms
between real and virtual corrections inside jets. It is a correlation
between the two jets. The resulting distribution factorizes
$$
\Sigma_{\ee}\!=S_{\qq}(\tau,\tin)\cdot C_{\qq}(\tau,\tin),
$$
with all factors depending on the SL variable
$$
\tau\!=\!\!\int_{\Eout}^{Q}\!\frac{dk_t}{k_t}\frac{N_c\as(k_t)}{\pi}
$$
The correlation function can be computed only in the planar
approximation (large $N_c$) by using the multi-dipole
formula\cite{BCM} for multi-soft-gluon distribution. Numerical studies
have been performed by Dasgupta and Salam\cite{DS} and the evolution
equation\cite{BMS} resumming all SL is
$$
\!\!\!\!\!\!\!\!\!\!\!\!
\partial_{\tau}\Sigma_{ab}\!=\!-(\partial_{\tau}R_{ab})\cdot\Sigma_{ab}
$$
$$
+\int_{\rm {  in}}\!\frac{d\Omega_k}{4\pi}\>\frac{(ab)}{(ak)(kb)}
\Big[\Sigma_{ak}\Sigma_{kb}-\Sigma_{ab}\Big]
$$
where $(ij)=1\!-\!\cos\theta_{ij}$ and
$$
R_{ab}=\tau\cdot \int_{\rm out}\!\frac{d\Omega_k}{4\pi}
\>\frac{(ab)}{(ak)(kb)}
$$
is the Sudakov radiator giving $S_{ab}\!=\!e^{-R_{ab}}$.  The
evolution equation involves general $ij$-dipoles (for $\ee$ one sets
$ij\Rightarrow\qq$). Since $k$ is inside the jet region, $ij$-dipoles
with $\theta_{ij}\!\sim\!\pi$ or $\theta_{ij}\!<\!\tin$ are coupled.

There are vary interesting properties for the correlation function
$C_{ab}$ for large $\tau$ (but experimentally accessible values are
$\tau\!<\!2\!-\!3$). The branching inside the jet region starts very
collinear to the primary quark with $\theta_{\rm
  br}\!<\!e^{-c\,\tau}$, there is a large buffer inside the jet
region\cite{DS}$^,$\cite{BMS}.  As a result one finds
$$
{ C_{\qq}(\tau,\tin)\sim e^{-\frac{c}{2}\,\tau^2}, \quad
  c=4.883\cdots}
$$
with $c$ universal (independent of $\tin$).
This shows that, at large $\tau$, the suppression in $C_{\qq}$ (due
to virtual corrections) is overwhelming the one in the Sudakov factor.

For $\theta_{ab}\!\ll\!1$ there is an amazing connection with small-x
dynamics. Indeed the value of $c$ is obtained by noticing that the
evolution equation for $\Sigma_{ab}$ is very similar to the Kovchegov
equation\cite{K} for the $S$-matrix even if the relevant multi-gluon
phase space regions are completely different (comparable angles in jet
physics while comparable transverse momenta in small-x physics).
Detailed studies of the connection between the two problems have been
performed\cite{MM}$^,$\cite{MO} for heavy quark pair production in
certain regions of phase space.

Non-global logs are difficult to compute.  Asymptotic estimates for
large $\tau$ are not physically relevant so that one needs to relay on
numerical computation\cite{DS}. Moreover its seems not easy to avoid
the large $N_c$ approximation. However, in some cases it is possible
to avoid/neglect non-global logs. This is the case when measurements
are in part of phase space but one is able to recover radiation in the
full phase space via momentum conservation (e.g. azimuthal
correlations). The other possibility is to work with a part of phase
space, but large enough. In particular, a rapidity region $Y$ in
hadron-hadron collision leads to non-global correction which are of
order $e^{-Y}$.



\begin{thebibliography}{99}
  
\bibitem{SigBet} S.Bethke, Nucl. Phys. C121 (03) 74
  [hep-ex/0211012].
  
\bibitem{DMW} Y.Dokshitzer, B.Webber, Phys. Lett. B352(1995)451
  [hep-ph/9504219]; Y.Dokshitzer, G.Marchesini, B.Webber, Nucl. Phys.
  B469 (1996) 93 [hep-ph/9512336]

\bibitem{KS} G.Korchemsky, G.Sterman, Nucl. Phys. B437(95)415
  [hep-ph/9411211]; Nucl. Phys. B555(1999)335-351 [hep-ph/9902341]

\bibitem{GR} E. Gardi, J. Rathsman, Nucl. Phys. B609(01)123 [hep-ph/0103217]

\bibitem {DM} M.Dinsdale, J.Maxwell, ICHEP04 contribution [hep-ph/0408114]

\bibitem{3jet-ee}
A. Banfi, Y.Dokshitzer, G.Marchesini, G.Zanderighi, JHEP 0105:040,2001
[hep-ph/0104162]; JHEP 0103:007,2001 [hep-ph/0101205];
Phys.Lett.B508:269-278,2001 [hep-ph/0010267]; JHEP 0007:002,2000
[hep-ph/0004027].
 
\bibitem{3jet-DIS}
A.Banfi, G.Marchesini, G.Smye, JHEP 0204:024,2002 [hep-ph/0203150];
A.Banfi, G.Marchesini, G.Smye, G. Zanderighi, JHEP 11:066,2001
[hep-ph/0111157]

\bibitem{3jet-hh}
A.Banfi, G.Marchesini, G.Smye, JHEP 08:047,2001 [hep-ph/0106278]

\bibitem{BS} J.Botts,~G.Sterman,~Nucl.Phys. B325:62,1989; N.Kidonakis,
  G.Sterman, Nucl.Phys.B505:321-348,1997 [hep-ph/9705234];
  N.Kidonakis, G.Oderda, G.Sterman, Nucl.Phys.B531:365-402,1998
  [hep-ph/9803241]

\bibitem{BSZ} A.Banfi, G.Salam and G.Zanderighi,
  Phys.Lett.B584:298-305,2004 [hep-ph/0304148]; JHEP 01(02)018
  [hep-ph/0112156]; JHEP 08:062,2004 [hep-ph/0407287]; and
  [hep-ph/0407286]

\bibitem{DS-DIS}
  M.Dasgupta,~G.Salam,~J.Phys.~G30: R143,2004~[hep-ph/0312283]; JHEP
  08:032,2002 [hep-ph/0208073]; Eur.Phys.J.C24:213-236,2002
  [hep-ph/0110213]; JHEP 02:001,2000 [hep-ph/9912488]
  
\bibitem{DS} M.Dasgupta,G.Salam, Phys.Lett.B512:323-330,2001
  [hep-ph/0104277]; JHEP 03:017,2002 [hep-ph/0203009];

\bibitem{BMS} A.Banfi, G.Marchesini and G.Smye, JHEP 08(02)006 [hep-ph/0206076]
  
\bibitem{BKS} C.Berger,T.K\'ucs,G.Sterman, Phys. Rev, D65:094031,2002
  [hep-ph/0303051]

\bibitem{YP}  Y.Dokshitzer, G.Marchesoini, JHEP 03:040,2003 [hep-ph/0303101]

\bibitem{BCM} A.Bassetto,M.Ciafaloni and G. Marchesini, Phys. Rep.100(1983)201

\bibitem{K} Y.Kovchegov,~Phys.Rev. D60:034008,1999 [hep-ph/9901281];
Phys.Rev.D61:074018,2000 [hep-ph/9905214];

\bibitem{MM} G.Marchesini and A.Mueller, Phys. Lett. B575(03)37 
[hep-ph/0308284]
\bibitem{MO} G.Marchesini and E.Onofri, JHEP 07(04)031 [hep-ph/0404242]

\end{thebibliography}
\end{document}